\documentclass[a4paper,twocolumn,showpacs,superscriptaddress]{revtex4}
\usepackage{graphicx}
\usepackage{amsmath,amssymb,braket}
\usepackage{hyperref}
\usepackage{color}

\newcommand{\hidden}[1]{}

\begin{document}
\preprint{MS \#L11-09414}

\title{Measurement of heavy-hole spin dephasing in (InGa)As quantum dots}
\author{R. Dahbashi}
\author{J. H{\"u}bner}
\author{F. Berski}
\author{J. Wiegand}
\affiliation{Institute for Solid State Physics, Leibniz
Universit{\"a}t Hannover, Appelstr. 2, D-30167 Hannover, Germany}
\author{X. Marie}
\affiliation{Universit\'e de Toulouse; INSA, UPS, CNRS; LPCNO, 135 avenue de Rangueil, F-31077 Toulouse, France}
\author{K. Pierz}
\author{H. W. Schumacher}
\affiliation{Physikalisch Technische Bundesanstalt, Bundesallee 100, D-38116 Braunschweig, Germany}
\author{M. Oestreich}
\affiliation{Institute for Solid State Physics, Leibniz
Universit{\"a}t Hannover, Appelstr. 2, D-30167 Hannover, Germany}

\date{\today}

\begin{abstract}
We measure the spin dephasing of holes localized in self-assembled (InGa)As quantum dots by spin noise spectroscopy. The localized holes show a distinct hyperfine interaction with the nuclear spin bath despite the \textit{p}-type symmetry of the valence band states. The experiments reveal a short spin relaxation time $\tau_{fast}^{hh}$ of 27 ns and a second, long spin relaxation time $\tau_{slow}^{hh}$ which exceeds the latter by more than one order of magnitude. The two times are attributed to heavy hole spins aligned perpendicular and parallel to the stochastic nuclear magnetic field. Intensity dependent measurements and numerical simulations reveal that the long relaxation time is still obscured by light absorption, despite low laser intensity and large detuning. Off-resonant light absorption causes a suppression of the spin noise signal due to the creation of a second hole entailing a vanishing hole spin polarization.
\end{abstract}


\maketitle

The spin of heavy-holes in quantum dots is carefully examined as a possible candidate for qubit implementation in solid-state based quantum information devices and has seen a large increase in research in recent years. Measurements on In(Ga)As/GaAs quantum dot (QD) ensembles yield hole spin dephasing times at zero external magnetic field of tens of nanoseconds.\cite{PhysRevLett.102.146601} Calculations show that these rather short hole spin relaxation times result from strain induced dipole-dipole interaction with the nuclei and that much longer heavy-hole (HH) spin relaxation times in Ising-type nuclear-spin interacting systems are finally limited by single hole-acoustic-phonon scattering and two-phonon processes.\cite{PhysRevB.78.155329} The strength of the hole-hyperfine interaction has been extracted from pump-probe and time-resolved photoluminescence and resonance fluorescence experiments whereat the pump-probe experiments demonstrate the quenching of the hole-spin relaxation by an external longitudinal magnetic field which exceeds the effective nuclear
field.\cite{PhysRevLett.102.146601,PhysRevLett.105.257402} Further experiments on single In(Ga)As/GaAs QDs show optical initialization of hole spins by optical pumping and highly coherent hole spins in coherent population trapping.\cite{Nature.451.441,Science.325.70}

All the above experiments are based on different kinds of setups with significant perturbations of the QD system. In this letter, we demonstrate spin noise spectroscopy (SNS) as an alternative low perturbation and in principle quantum mechanical non-demolition experiment \cite{PhysRevLett.95.216603,PhysRevLett.101.206601,Physica_E.43.569} to study the interaction of localized heavy holes with the surrounding nuclear spin bath in dependence on external longitudinal and transverse magnetic fields. Spin noise spectroscopy has been used on QD ensembles once before, however, mainly to study the anisotropy of the heavy hole g-factor.\cite{PhysRevLett.104.036601}
In the following we apply magnetic fields at oblique angle with respect to the  projection direction ($z$) of our all optical spin sensitive probing technique. The effective magnetic field given by the superposition of the stochastic nuclear and the applied external magnetic field divides the spin dynamics into a longitudinal and transverse component. The projection of both components onto the detection direction yields simultaneous access to the complete spin dynamic.

\begin{figure}[htb]
  \centering
  \includegraphics[width=0.75 \columnwidth]{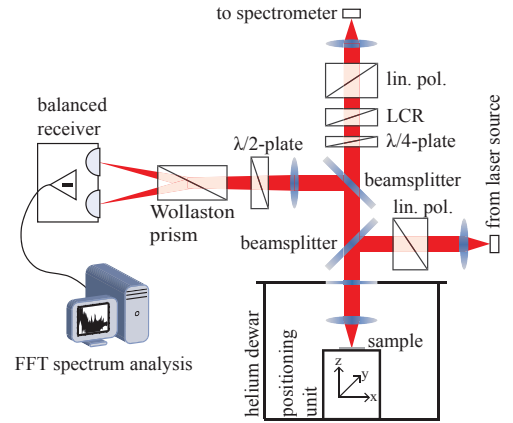}
  \caption{(Color online) Experimental setup.
  }
\label{fig:Aufbau}
\end{figure}

Spin noise spectroscopy utilizes in general the statistical fluctuations of a spin polarization at thermal equilibrium which contain according to the fluctuation-dissipation theorem the full temporal dynamics of the spin system under infinitesimal external perturbation. The spin fluctuation is recorded via non-resonant Faraday rotation of a linearly polarized probe laser. Figure~\ref{fig:Aufbau} shows the corresponding experimental setup. The light source is a commercial low noise, tunable diode laser in Littman configuration. The linearly polarized laser light is focused to a beam waist of about $1~\mu$m on the sample which is mounted in a free beam confocal microscope in a liquid Helium dewar at a constant temperature of 4.2~K. The sample is a single layer of InAs/GaAs quantum dots grown by molecular beam epitaxy on (100)-oriented GaAs inside the antinode of a $\lambda$-Bragg cavity with 13 and 30 GaAs/AlAs layers for the top and bottom mirror, respectively. The high finesse microcavity enables SNS measurements in reflection and enhances the Faraday rotation noise signal without increasing the optical shot noise. The QD density varies spatially from zero to about 100 dots/$\mu\text{m}^2$ and some of these QDs are filled by a single hole due to the MBE p-type background of typically $10^{14}$ holes/cm$^{3}$. We choose for our SNS measurement a sample region with high dot density to increase the signal to noise ratio and to ensure that each QD is occupied in average by less than one hole from the background doping. Magnetic fields of up to 30~mT are applied both in Faraday and Voigt geometry. The spin induced stochastic Faraday rotation of the linear polarization of the reflected laser light is resolved outside the He dewar by a combination of a Wollaston prism and an 80~MHz low noise balanced photo receiver. The detected electrical signal is amplified by a low noise amplifier, digitized with 180~MHz sampling rate in the time domain, and Fourier transformed in real time. The noise background due to optical shot noise of the laser and electrical noise of the balanced receiver and the amplifier is eliminated by subtracting spin noise spectra with and without applied external magnetic field from each other.\cite{Physica_E.43.569} This method works well due to the strong influence of the magnetic field on the spin noise spectra. 

\begin{figure}[tb]
  \centering
  \includegraphics[width=1 \columnwidth]{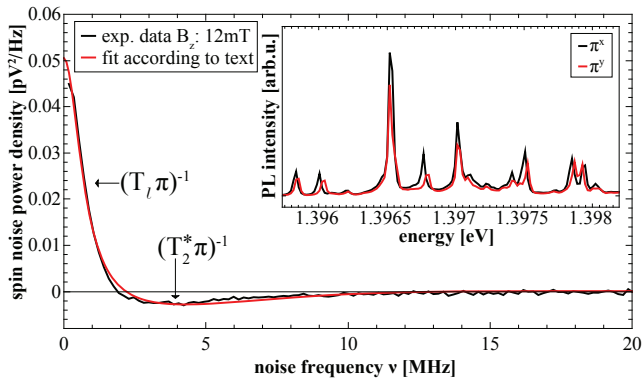}
  \caption{(Color online) 
  Typical spin noise difference spectrum (black line) at an excitation
  intensity of 140~$\mu\text{W}/\mu\text{m}^2$ and the corresponding
  fit (red curve) with one Lorentzian and one Gaussian curve centered at zero frequency.
  (Inset) Polarization resolved QD PL spectrum measured 
  in growth direction.  
  }
\label{fig:Musterkurve_PL}
\end{figure}

First, we characterize the sample by polarization resolved photoluminescence (PL) measurements.
The inset of Fig.~\ref{fig:Musterkurve_PL} shows the measured PL for excitation and
detection in growth direction. The measured PL line width 
of individual QDs of 40~$\mu$eV is limited by the 
resolution of the spectrometer. Statistics over many sample
spots show that about half of the measured QDs do not show
an anisotropic exchange interaction splitting which is a good indication that these
QDs are charged.\cite{PhysRevB.65.195315} We observe QD PL in the growth direction only within the
spectral width of the cavity resonance of 2.4~meV. Measurements from the
side of the sample with a larger laser spot yield a QD PL width 
of 30~meV and we estimate accordingly
that about 50 charged QDs are located within our 1~$\mu$m 
laser spot. 

Next, we measure spin noise (SN) at the identical sample position as the 
depicted micro PL. 
Figure~\ref{fig:Musterkurve_PL} shows an exemplary SN 
spectrum at $B_z^{ext}=12$~mT minus a SN spectrum at $B_z^{ext}=0$~mT where $z$ is the growth
direction and the direction of laser propagation. The laser wavelength
has been red-detuned for this SN measurement from the closest charged QD resonance 
by approximately 20 times the full width 
at half maximum (FWHM) of the QD ground state transition.
We do not observe a significant dependence
of the SN spectrum on laser detuning which indicates that we indeed measure the 
spatio-spectral average of the QD ensemble in the laser focus. 

The SN spectrum in Fig.~\ref{fig:Musterkurve_PL} is fitted by one Lorentzian and one Gaussian 
curve, both centered at zero frequency. The fitting analysis
reveals that the areas of the two curves are in very good 
approximation equal but have opposite signs. This is reasonable 
since the integrated spin noise power does not depend on magnetic field
and the opposite sign results from calculating a difference spectrum. 
The fitting analysis yields a
spectral $\nu_\text{FWHM}$ of 1.5~MHz for the Lorentzian curve ($B_z^{ext}=12$~mT) and 11.9~MHz
for the Gaussian curve ($B_z^{ext}=0$~mT). The two $\nu_\text{FWHM}$ translate by
$\nu_\text{FWHM} = 1/(\pi\tau^{hh}_{fast,slow})$ \cite{11} to hole spin relaxation times $\tau^{hh}_{slow}=215$~ns and $\tau^{hh}_{fast}=27$~ns, respectively.
The spin relaxation time of $27$~ns 
results from the transverse stochastic
nuclear spin orientation in the QDs. The statistical nuclear spin fluctuations yield for the holes an effective nuclear 
field $B^N_{hh}$ due to dipole-dipole interaction and $\tau^{hh}_{fast}$ is the corresponding QD HH spin
transverse $T_2^\ast$ time, i.e., describes the spin dynamics of the HH spin components
which are perpendicular to $B^N_{hh}$. The $T_2^\ast$ time translates with a HH g-factor of 0.15 \cite{PhysRevLett.104.036601}
to $B^N_{hh}\approx 6$~mT   
which is in excellent agreement with pump-probe experiments on very similar ensembles of p-doped InAs-QDs 
and theory.\cite{PhysRevLett.102.146601} 
The excellent
agreement confirms that our QDs are in fact occupied by heavy-holes.
We can exclude loading of the QD with two holes since spin noise vanishes 
if both, the HH spin-up and spin-down, states are occupied.
We can also exclude loading by a single electron since electron spin relaxation
times in InAs-QDs are shorter by one order of magnitude at $B^{ext}_z=0$~mT due
to the much more efficient hyperfine interaction. 

\begin{figure}[tb]
  \centering
  \includegraphics[width=\columnwidth]{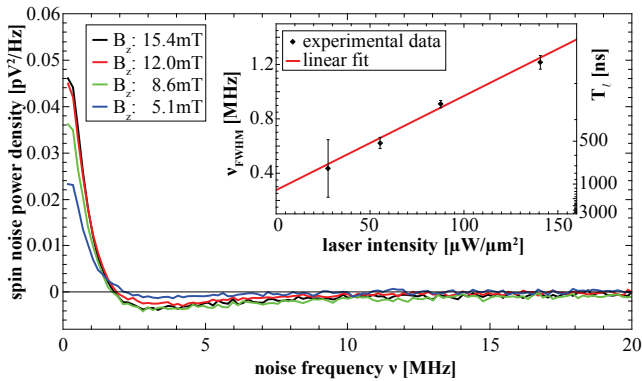}
  \caption{(Color online) Spin noise difference spectra at $B^{ext}_z=5.1$~mT 
  (blue), 8.6~mT (green), 12~mT (red), and 15.4~mT (black) minus $B^{ext}_z=0$~mT.
  (Inset) Intensity dependence of the narrow SN peak ($1/(\pi\tau_{slow}^{hh})$) 
  for $B^{ext}_z=30$~mT.
  }
\label{fig:fig03}
\end{figure}

Figure~\ref{fig:fig03} depicts the dependence of the SN spectrum on 
a longitudinal magnetic field in more detail. The measurement 
clearly shows that the spectrally broad
$B_z^{ext}=0$~mT, $T_2^\ast$ spin noise power is gradually transferred with
increasing $B^{ext}_z$ into the narrow, longitudinal SN peak. 
This transfer of spin noise power and the width and amplitude of the narrow peak saturate for
$B^{ext}_z \gg B^N_{hh}$. 
 
\begin{figure}[tb]
  \centering 
  \includegraphics[width=\columnwidth]{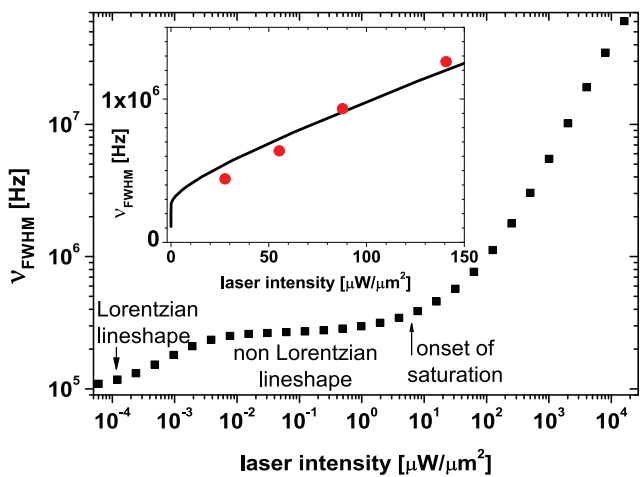}
  \caption{(Color online) Calculated FWHM of the SN power in dependence on laser intensity
  for a hypothetic $\gamma_{I=0}=10^5$~Hz ($\cong \tau_{slow}^{hh}\approx 31~\mu$s ) at zero laser intensity.
  (Inset) Comparison of calculated (black line) and measured (red dots) FWHM. The
  probability of light absorption by a single QD at resonant excitation is not exactly 
  known in our microcavity sample and set as
  adjustable parameter to $1/6\cdot 10^{-3}$. 
  }
\label{fig:fig04}
\end{figure}

We have also studied at $B^{ext}_z=0$~mT the spin relaxation time of HH spins pointing along the stochastic 
nuclear magnetic field 
by measuring a SN spectrum at $B^{ext}_{x,z}=0$~mT minus a SN spectrum at a
transverse external magnetic field $B^{ext}_x$. A high transverse magnetic field \cite{12} of
$B^{ext}_x=18$~mT shifts the complete SN power out of the detection window, i.e., the difference spectrum
only includes the SN power at $B^{ext}_{x,z}=0$~mT. The broad $T_2^\ast$ SN spectrum is 
in this configuration less clearly visible
since the narrow, longitudinal ($T_l$) and the broad transverse ($T_2^\ast$) SN peaks have the 
same sign. For $B_x^{ext}=6$~mT, we measure only 
about half the amplitude of the longitudinal SN at $B^{ext}_{x,z}=0$~mT
since the total magnetic field is tilted in average by $45^\circ$ with respect to the
z-axis if $B_x^{ext}=B^N_z$. The tilting reduces the observed $T_l$ spin noise power 
by a factor of two since the projection of the HH spins pointing
along the tilted magnetic field on the z-direction is $1/\sqrt{2}$. 
The $T_l$ SN measurements with transverse magnetic fields thereby confirm the
amplitude of $B^N$ which has been extracted from $T^\ast_2$ measured with longitudinal magnetic fields.
We want to point out that a tilted magnetic field reveals in SNS 
in general $T_l$ and $T_2^\ast$ at the same time, the first
as a SN spectrum at $\nu=0$ and the second shifted by the Larmor frequency. 

Next, we want to demonstrate that the measured $\tau^{hh}_{slow}$ is not the intrinsic
longitudinal spin relaxation time but significantly altered by 
light absorption.
Spin noise spectroscopy can be in principle a non-demolition technique but 
our QD measurements reveal a clear intensity dependence for $\tau^{hh}_{slow}$.
The inset in Fig.~\ref{fig:fig03} shows that the FWHM of the narrow peak at $B_z^{ext} = 30$~mT decreases almost linearly 
with decreasing excitation intensity.
A linear extrapolation to zero 
intensity might be tempting but the intensity dependent broadening of the SN spectrum is different
for each QD due to the different detunings of the laser frequency from the individual QD resonances.
The dependence of the integrated SN spectrum of the QD ensemble on the intensity $I$
can be calculated by
\begin{equation}\label{equ:broadening}
  \int \limits_{0}^{+\infty} G(E,I)\cdot n^2(E)\cdot R(E,I)\cdot L(\gamma(E,I)) dE
\end{equation}
where $G(E)$ is the Gaussian energy distribution of the QDs ground state resonance
with a measured FWHM 
of 30~meV, $n$ is the dispersive part of the QD refractive index at the laser
frequency, $R(E,I)=1-\alpha(E,I)\tau_{PL}$ 
includes that optically excited QDs with the radiative lifetime
$\tau_{PL}$ do not contribute
to the HH SN spectrum due to Pauli blockade of the heavy holes, and $L$ is the Lorentzian 
SN spectrum of a single QD with a FWHM
$\gamma(E,I) = \gamma_{I=0} + \alpha(E,I)/\pi$ where $\gamma_{I=0}=10^5$~Hz and $\alpha(E,I)$ is the number of 
absorbed photons per second. The calculated $\nu_\text{FWHM}$ is depicted in Fig.~\ref{fig:fig04} 
and reveals three distinct
intensity regimes. A moderate increase
of $\nu_\text{FWHM}$ at extremely low laser intensities due to excitation of nearly resonant
QDs, a weak increase at intermediate 
intensities, and a strong increase for very high intensities where all QDs are strongly
excited. A comparison with our experimental data (red dots in the inset of Fig.~\ref{fig:fig04}) 
shows, firstly, the good agreement between calculations and experiment, secondly, that the
laser intensities of our experiment coincide with the high intensity limit from which a linear
extrapolation to zero intensity is not justified, and, thirdly, that the intensity 
dependent broadening is significant for $T_l$ but not for $T_2^\ast$. At first sight, the calculations
might imply that $T_l$ measurements on single QDs by SN are experimentally impractical, 
however, the relative SN power around $\nu=0$ increases significantly with decreasing 
laser intensity since quasi-resonant QDs are only weakly perturbed and thereby contribute 
strongly.\cite{13}  

In conclusion, we have measured in a quantum dot ensemble the $T_l$ and $T_2^\ast$ 
heavy hole spin relaxation times by spin noise spectroscopy. The $T_2^\ast$ time 
and the transverse magnetic field dependence both yield a heavy hole hyperfine interaction
with an effective stochastic nuclear magnetic field $B_N\approx 6$~mT. Intensity 
dependent measurements and calculations show that the very long $T_l$ heavy hole spin relaxation time
is significantly influenced by laser excitation and that a linear extrapolation to zero
intensity is not justifiable. 

We gratefully acknowledge the excellent technical support by R. H\"uther and financial support by the BMBF joint research project {\it QuaHL-Rep}, 
the Deutsche Forschungsgemeinschaft in the framework 
of the priority program "SPP 1285 - Semiconductor Spintronics" and the excellence cluster "QUEST - 
Center for Quantum Engineering and Space-Time Research".


\begin{thebibliography}{13}
\bibitem{11}
The equation $\nu_\text{FWHM}=1/(\pi\tau)$ is strictly valid only for a Lorentzian line shape.
\bibitem{12}
Note that $\mu_BB\ll k_BT$. Thereby, magnetic field induced hole spin polarization is negligible small.
\bibitem{13}
At very low laser intensities, the electrical noise of our 80~MHz New Focus balanced receiver becomes much larger than the optical shot noise, i.e., an optimized low noise (low frequency) balanced receiver would be advantageous for single QD SN spectroscopy.
\end{thebibliography}
\end{document}